\documentclass[manuscript,nonacm]{acmart}

\usepackage{pgfplots}
\pgfplotsset{compat=1.17}
\usepackage{amsmath}
\usepackage{booktabs}
\usepackage{makecell}
\usepackage{multirow}
\usepackage{subcaption}
\usepackage{longtable}
\AtBeginDocument{%
  }

\setcopyright{acmlicensed}
\copyrightyear{2018}
\acmYear{2018}
\acmDOI{XXXXXXX.XXXXXXX}

\acmConference[Conference acronym 'XX]{Make sure to enter the correct
  conference title from your rights confirmation emai}{June 03--05,
  2018}{Woodstock, NY}
\acmISBN{978-1-4503-XXXX-X/18/06}

\begin{document}

\title{Anger Speaks Louder? Exploring the Effects of AI Nonverbal Emotional Cues on Human Decision Certainty in Moral Dilemmas}


\author{Chenyi Zhang}
\authornote{Both authors contributed equally to this research.}
\email{chenyiooo12@outlook.com}
\author{Zhenhao Zhang}
\authornotemark[1]
\affiliation{%
  \institution{Southern University of Science and Technology}
  \city{Shenzhen}
  \state{Guangdong}
  \country{China}
}

\author{Wei Zhang}
\affiliation{%
  \institution{Shenzhen University}
  \city{Shenzhen}
  \state{Guangdong}
  \country{China}
  }

\author{Tian Zeng}
\affiliation{%
  \institution{Shenzhen University}
  \city{Shenzhen}
  \state{Guangdong}
  \country{China}
}

\author{Black Sun}
\affiliation{%
 \institution{Aarhus University}
 \city{Aarhus}
 \country{Denmark}
 }

\author{Jian Zhao}
\affiliation{%
  \institution{University of Waterloo}
  \city{Waterloo}
  \state{Ontario}
  \country{Canada}}

\author{Pengcheng An}
\authornote{Corresponding author.}
\email{anpc@sustech.edu.cn}
\affiliation{%
  \institution{Southern University of Science and Technology}
  \city{Shenzhen}
  \state{Guangdong}
  \country{China}
}

\renewcommand{\shortauthors}{Chenyi Zhang et al.}

\begin{abstract}
Exploring moral dilemmas allows individuals to navigate moral complexity, where a reversal in decision certainty—shifting toward the opposite of one's initial choice—could reflect open-mindedness and less rigidity. This study probes how nonverbal emotional cues from conversational agents could influence decision certainty in moral dilemmas. While existing research heavily focused on verbal aspects of human-agent interaction, we investigated the impact of agents expressing anger and sadness towards the moral situations through animated chat balloons. We compared these with a baseline where agents offered same responses without nonverbal cues. Results show that agents displaying anger significantly caused reversal shifts in decision certainty. The interaction between participant gender and agents’ nonverbal emotional cues significantly affects participants’ perception of AI’s influence. These findings reveal that even subtly altering agents’ nonverbal cues may impact human moral decisions, presenting both opportunities to leverage these effects for positive outcomes and ethical risks for future human-AI systems.
\end{abstract}

\ccsdesc[300]{Human-centered computing~Empirical studies in HCI}
\ccsdesc[300]{Computing methodologies~ Artificial
intelligence.}

\keywords{AI-Assisted Decision-making, Human-AI Collaboration, Emotion, Moral Dilemmas}


\maketitle

\begin{figure*}[h]   
    \centering
    \includegraphics[width=1\linewidth]{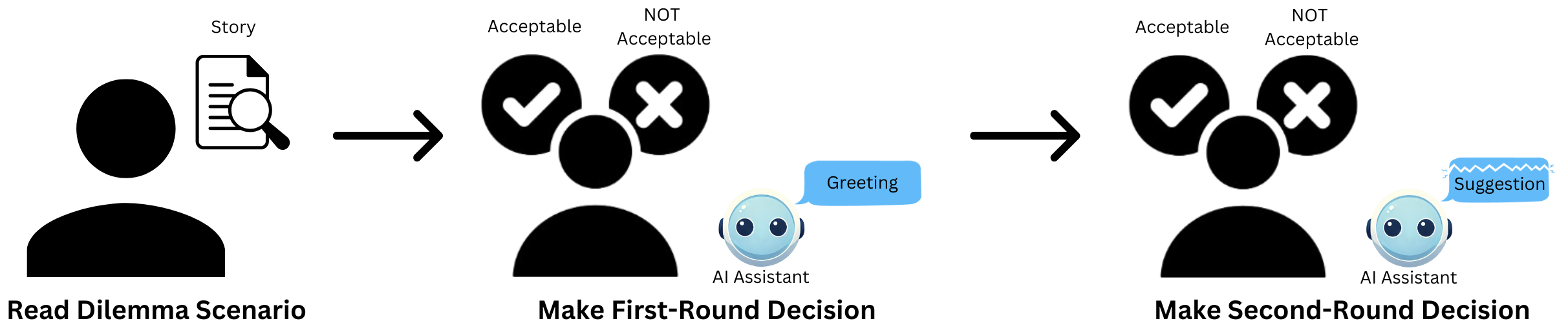} 
    \caption{Task Process: Each user is asked to read the scenario and make decisions in two rounds with an AI assistant. In the second round, the AI assistant provides suggestions about the dilemma wrapped in an emotional bubble. }
    \label{fig:process} 
\end{figure*}


\section{INTRODUCTION}
A moral dilemma arises when an individual must choose between two conflicting moral principles, where any decision will likely violate an important ethical standard. For instance, the famous Trolley Problem illustrates a scenario where a runaway trolley is heading towards five people tied to a track. You can pull a lever to divert the trolley onto another track, where only one person is tied. This situation forces you to choose between saving five lives by sacrificing one or not intervening to avoid directly causing harm \cite{thomson1984trolley}. Other true moral dilemmas include a doctor deciding how to allocate limited medical resources and military leaders choosing targets to minimize civilian casualties during warfare. These scenarios demonstrate the complexity and challenges inherent in moral dilemmas. In contrast, a situation where one must choose between two equally appealing job offers, while challenging, does not constitute a moral dilemma as it lacks the element of conflicting ethical principles.
Moral dilemma plays an important role in various fields. Moral dilemmas provide a framework for moral reasoning and promote moral development\cite{kohlberg1971stages, kohlberg1987psychology}. Moral dilemmas are effective tools for studying how different factors shape moral decision-making\cite{greene2008cognitive, christensen2014moral}. Moral dilemmas also provide an open environment for individuals to explore and creatively imagine potential solutions, enabling research and training in moral imagination\cite{johnson2014moral}.

When faced with complex ethical choices, unlike most decision-making tasks, differences in moral dilemma judgments are primarily influenced by moral traits, which refer to enduring characteristics such as empathy, fairness, and honesty that guide an individual's ethical reasoning. These traits are relatively stable and unlikely to change due to external factors in the short term \cite{luke2022temporal}. On the other hand, decision certainty plays a critical role in moral dilemmas. Decision certainty refers to the confidence an individual has when making a choice. Interestingly, uncertainty can sometimes reflect positive traits, as individuals experiencing uncertainty tend to engage in more systematic analysis \cite{pelham1995waxing,weary1997causal} and invest greater effort in processing information \cite{chaiken1989heuristic}.

Despite these insights, current research rarely regulates decision certainty in moral dilemmas, especially using Human Computer Interaction(HCI) design approaches. Most existing studies focus on understanding the factors influencing moral judgments\cite{schaich2006consequences,arutyunova2016sociocultural,klenk2022influence} or the outcomes of moral decisions\cite{detert2008moral,jeurissen2014tms}, but few concentrate on moral decision certainty. Based on this, we aim to find suitable HCI methods to actively influence human certainty in moral decision-making, encouraging them to be open to different moral principles. 


Artificial Intelligence(AI), especially with the development of Large Language Models(LLMs), has become a powerful tool for assisting in moral decisions. LLMs allow human decision-makers to explore various scenarios and potential outcomes before deciding. For example, Krügel et al. designed an experiment where ChatGPT provided moral advice to users in moral dilemmas and found that ChatGPT readily dispenses moral advice although it lacks a firm moral stance\cite{krugel2023chatgpt}. There have also been efforts to develop AI agents capable of autonomous decision-making in moral dilemmas \cite{zhang2023moral} or simulating discussions from multiple perspectives \cite{taubenfeld2024systematic}.

However, current AI assistants in moral dilemmas still heavily rely on verbal responses while overlooking the potential of non-verbal cues. Moral emotions theory suggests that people experience specific emotions in response to morally relevant issues in particular contexts \cite{tangney2007moral}. Emotions can convey complex information quickly and effectively, making them a powerful tool in guiding decision-making processes. Inspired by moral emotions theory, this study focuses on emotions as the non-verbal cue to be examined. 

We utilized AniBalloons, a set of chat balloon animations conveying Joy, Anger, Sadness, Surprise, Fear, and Calmness, as our nonverbal emotional cues\cite{an2023affective,an2024aniballoons}. AniBalloons are designed based on text bubbles, essential design elements for every conversational AI. Unlike other nonverbal emotional cues, they do not involve a specific race or gender and can be adapted to a conversational AI and Avatars. They have been proven in previous studies to be more inclusive and universal, and their expressions of specific emotions can be correctly perceived, making it easier for us to manipulate\cite{an2024aniballoons}. Additionally, they have been shown to effectively communicate the intended emotions without relying on the context of the message \cite{an2023affective,an2024aniballoons}.

Moral emotions theory suggests that people experience specific emotions in response to morally relevant issues in particular contexts\cite{tangney2007moral}. Among other-focused moral emotions, anger and sympathy are commonly observed. The anger caused by the perpetrator’s violating moral standards can be felt upon witnessing morally repulsive behavior aimed at others. Sympathy is central to the human moral affective system\cite{eisenberg2004social, eisenberg2006handbook}, often felt when seeing others in distress. Sympathy is usually shown as sadness\cite{barankova2019analysis}. Since our study aims to preliminarily explore the impact of emotions as nonverbal cues, we selected two emotions, anger and sadness, from the six basic emotions to assign to the AI assistant.

Our research questions are as follows:
\begin{itemize}
\item{\textbf{RQ1}}: How do the AI assistant's nonverbal emotional cues influence human decision certainty in moral dilemmas?
\item{\textbf{RQ2}}: How do humans perceive the AI assistant's influence?
\end{itemize}

In this paper, we present the findings of an online user study by investigating the impact of AI nonverbal emotional cues on human decision certainty in moral dilemmas. We developed an interface where participants could interact with an AI assistant whose verbal responses are generated by LLM but without emotion, make decisions in moral dilemmas, and rate their certainty. We compared AI assistants expressing anger and sadness via different AniBalloons, with a baseline condition where the agents provided identical responses without nonverbal cues. The experimental data indicate that angry cues, compared with sadness cues, could significantly alter people's certainty, and the perceived influence of the two cues shows considerable gender differences in statistics.

\section{RELATED WORK}
\subsection{Moral Dilemmas and Decision Certainty}

Moral dilemmas present situations where individuals must choose between conflicting moral principles, offering valuable insights into ethical reasoning and decision-making \cite{marcus1980moral}. Moral dilemma tasks have been a much-appreciated experimental paradigm in empirical studies on moral cognition for decades \cite{christensen2012moral}. Experimental philosophers take them to test and challenge the reliability of moral intuitions \cite{bagnoli2011morality}. These tasks investigate how people make moral decisions and the cognitive processes involved \cite{greene2001fmri}, and differentiate between emotional and rational components in moral judgment \cite{greene2008cognitive}. Additionally, moral dilemmas are employed to study the impact of empathy on moral decisions \cite{gleichgerrcht2013low}.

Decision certainty refers to the level of confidence an individual has in the correctness or appropriateness of their decision. High levels of uncertainty in decision-making can be advantageous. When people are uncertain about their self-views, they tend to engage in more systematic processing \cite{pelham1995waxing,weary1997causal}. According to Chaiken et al.'s sufficiency threshold hypothesis, individuals will exert whatever effort is necessary to achieve a sufficiently confident assessment of message validity. When their actual confidence or certainty is below the desired level, they will invest more effort in processing. This is because feeling certain acts as an internal cue that one is already correct and accurate, suggesting that further processing might not be necessary \cite{chaiken1989heuristic}. Additionally, a more questioning attitude and openness to information gathering are essential. Higher critical thinking skills and hesitancy in decision-making could be encouraged in questioning attitude\cite{hoffman2004relationship}. Thus, high uncertainty could drive more thorough and critical information processing, leading to better decision-making outcomes.

\subsection{Decision Making with AI}
AI has become increasingly utilized to aid decision-making across various fields thanks to significant technological advancements, such as criminal justice \cite{doswell2004weather, dressel2018accuracy}, admissions \cite{cheng2019explaining}, financial investment \cite{green2019principles}, and medical diagnosis \cite{cai2019hello}, among others. Concerns regarding AI’s accuracy, safety, ethics, and accountability \cite{binns2018s, cai2019hello} have driven the widespread adoption of AI-assisted decision-making in practical applications \cite{wang2004communicating, zhang2020effect}.

Since the emergence of LLMs, an increasing number of researchers have chosen to design AI systems based on these models. Eigner and Händler comprehensively analyze the factors influencing decision-making with LLM support, highlighting the interplay of technological, psychological, and decision-specific determinants \cite{eigner2024determinants}. Yang et al. examine the voting behaviors and biases of LLMs like GPT-4 and LLaMA-2, finding that voting methods, presentation order, and persona variations affect their alignment with human voting patterns \cite{yang2024llm}. Similarly, Chen et al. and Liu et al. investigated these biases and emphasize the need to carefully integrate LLMs into democratic processes due to potential biases and reduced preference diversity \cite{chen2024humans, liu2024make}.

Leveraging LLMs, human-AI moral decision-making is a complex and increasingly relevant field that raises important questions about accountability, transparency, and bias. For instance, Greene et al. investigated the neural basis of moral decision-making using fMRI, highlighting the role of emotional and cognitive processes in ethical judgments \cite{greene2001fmri}. Malle et al. explored how people perceive and trust AI systems in moral decisions, emphasizing the importance of aligning AI behavior with human moral values to ensure acceptance and reliability \cite{malle2015sacrifice}. These works underscore the need for interdisciplinary approaches to develop AI systems that can make morally sound decisions while being transparent and accountable to human users.

\subsection{Nonverbal Emotional Cues of AI}

Emotions constitute significant impacts in interpersonal decision making\cite{lerner2015emotion}. Expressions of anger prompt concessions from negotiation partners\cite{van2004interpersonal} and more cooperative strategies in bargaining games\cite{wubben2009emotion}. Rind and Bordia found that Communicating gratitude triggers others' generosity\cite{rind1995effect}. What's more, communicating disappointment with a proposal can evoke guilt in a bargaining partner and motivate reparative action\cite{lelieveld2013does}. 
Marreiros, G., et al. demonstrates that emotional agents simulating human-like emotional dynamics can affect argument selection and evaluation, aiding in reaching consensus in simulations\cite{marreiros2005emotion}.

Since the development of AI systems, the emotional expression methods of AI systems have become increasingly diverse, which can be categorized based on different target modalities, including verbal and nonverbal approaches. However, text has inherent limitations in conveying emotions\cite{aoki2022emoballoon,an2022vibemoji}. Nonverbal cues like facial expressions, vocal features, and bodily motions can convey more nuanced emotions\cite{vinciarelli2009social}, which are absent in pure text communication. Currently, some studies have explored nonverbal emotional cues, focusing on various target modalities.

Motion-based approaches leverage body and skeletal joint movements to enable robots to express emotions. Research in this area has utilized advanced machine learning techniques to achieve human-like emotional expressions. For instance, Nishimura et al. employed Generative Adversarial Networks (GANs) to generate human-like motion patterns that mimic real interactions \cite{nishimura2020human}. Tuyen et al. focused on extracting and clustering joint movement descriptors, allowing robots to adapt their interaction patterns over time based on individual users' habitual behaviors \cite{tuyen2020learning}. Similarly, Suguitan et al. utilized a puppeteer robot to display emotional movements learned through variational autoencoders, effectively generalizing affective motion generation \cite{suguitan2020moveae}.

Visual nonverbal expression has also been a critical area of exploration. Ke et al. combined bimodal emotional recognition with fuzzy decision-making models to provide human-like emotional feedback, effectively conveying nonverbal emotional cues \cite{ke2020interactive}. Zheng et al. investigated the use of touch as a nonverbal emotional cue, demonstrating that the length and type of touch significantly influence the perceived strength and naturalness of emotions \cite{zheng2019kinds}.

In conversational AI, nonverbal emotional cues have been studied through innovative visual and textual methods. Choi and Aizawa proposed a system that uses typefaces to communicate emotions \cite{choi2019emotype}. Wang et al. designed a chat system with animated dynamic text that conveys users’ emotional states \cite{wang2004communicating}. Additionally, Liu et al. explored the use of expressive biosignals in chat systems to reflect social dynamics and enhance emotional communication \cite{liu2017supporting}.

Recent research has explored chat balloons as an innovative medium for visually expressing emotions. Aoki et al. introduced EmoBalloon, which uses the "explosion" shape of chat balloons to depict emotional intensity \cite{aoki2022emoballoon}. Chen et al. examined the use of chat balloon colors to convey the emotional tone of voice messages, such as excitement, anger, sadness, and calmness \cite{chen2021bubble}.

To integrate chat balloons with animations for emotional expression, An et al. developed AniBalloons, a set of chat balloon animations designed to communicate six basic emotions—Joy, Anger, Sadness, Surprise, Fear, and Calmness—based on Ekman’s basic emotion theory \cite{ekman1992there,ekman1992argument}. AniBalloons successfully conveyed emotions even without contextual message content and enhanced human-chatbot interaction by serving as animated teasers on message bubbles \cite{an2023affective,an2024aniballoons,an2024emowear}.

Inspired by these works, we adopted AniBalloons as the nonverbal emotional cue in the design of our AI assistant. 

\section{METHOD}
\subsection{Experimental Design}
\begin{figure*}[ht]
    \centering
    \includegraphics[width=\textwidth]{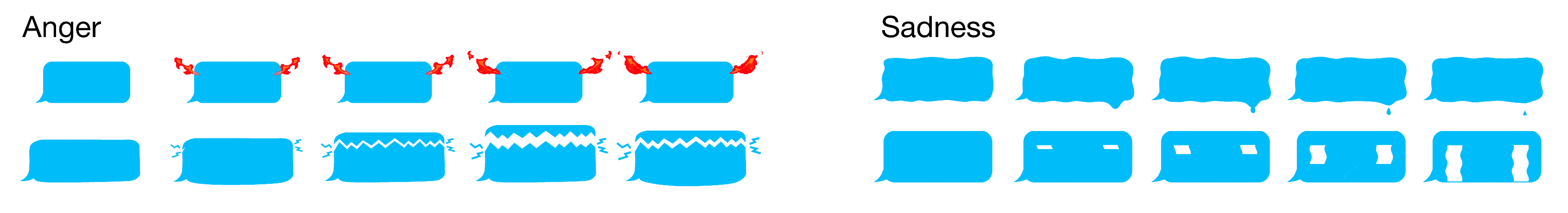}
    \caption{AniBalloons used in this experiment. The left-side column features two animations representing anger, while the right-side column features two animations representing sadness.}
    \label{fig:bubbles}
\end{figure*}

We selected two emotions, including anger and sadness to assign to the AI assistant. Correspondingly, we selected AniBalloons that express anger and sadness. Figure \ref{fig:bubbles} listed the animation style we took from AniBalloons to express anger and sadness.

Our study employs a mixed design, dividing the participants into three groups. 
\begin{itemize}
    \item \textbf{Baseline}: The AI assistant expressed opinions solely through verbal responses generated by a Large Language Model.
    \item \textbf{AC} (Anger Cues): The AI assistant expressed the same verbal responses while also conveying emotions of anger via AniBalloons.
    \item \textbf{SC} (Sadness Cues): The AI assistant expressed the same verbal responses while also conveying emotions of sadness via AniBalloons.
\end{itemize}

Participants in each group made moral dilemma decisions in four different scenarios, the order of which was balanced.

\subsection{Materials}

In this study, the moral dilemma scenarios were selected from the 48 stories proposed by Körner under the CNI model theory\cite{korner2020using}.
The CNI model\cite{gawronski2017consequences} allows researchers to quantify sensitivity to consequences (C), sensitivity to moral norms (N), and the general preference for inaction versus action irrespective of consequences and norms (I) in responses to moral dilemmas. 
The 48 scenarios are widely used in psychological research related to morality\cite{luke2022temporal,korner2020using,ng2024thinking,gawronski2022moral}.


Regarding the design of an AI assistant, in most existing AI-assisted decision-making work, AI typically plays the role of offering suggestions or providing additional information. We have specifically designed an AI assistant that always presents opposing viewpoints. This behavior pattern has been proven to have a positive effect on joint decision-making.
The presentation of opposing viewpoints can encourage deeper reflection on a dilemma, a concept supported by various studies. Diehl and Stroebe highlight that when someone role-plays a position that critiques the favored alternative, it can prevent premature closure of the decision-making process and presumably increase the consideration of alternative options\cite{diehl1987productivity}. Schwenk and Schweiger has found that devil's advocate, which means a person who intentionally takes an opposing argues against a prevailing idea to provoke discussion, have some beneficial consequences for promoting and managing cognitive conflict on decision making\cite{schwenk1990effects, schweiger1986group}. Based on these studies, our AI assistant are designed to play the ``devil", speaking against the participant's choice.

The AI assistant's text response to participants' decisions in moral dilemmas were generated using the ChatGPT-4o model. In the prompt provided to ChatGPT, the consequences of the choice made by the participant in the first round will be analyzed, including the perpetrator's acts of violence and the tragic experiences suffered by the victim(s). It is important to note that the AI assistant's text responses are consistent across the three groups and without emotion. We must ensure that the nonverbal emotional cues perceived by the participants in AC and SC come only from AniBalloons. The model generated a 30-35 word response, which was then displayed by the AI assistant on the interface. 
The prompts and generated responses we applied are shown in Table \ref{tab:opinions}.

To align with the behavior pattern of the AI assistant, our criteria for selecting stories required clearly defined perpetrators and victims, allowing the AI assistant to express anger toward perpetrators and sadness toward victims reasonably. Based on this, we chose two scenarios. The first involves deciding between paying a ransom to a guerrilla group, which enables further violence, or refusing and resulting in a hostage's death. The second involves choosing between betraying a colleague, leading to their execution by militants, or not betraying them, which results in the death of a busload of strangers. Each story has two versions: one where a proscriptive norm prohibits action with benefits outweighing the costs, and the other prescribes action with benefits smaller than the costs. The scenarios are shown in Table \ref{tab:scenarios}. The four scenarios appeared randomly for each participant during the task. 

\subsection{Procedure}

\begin{figure}[h]   
    \centering
    \includegraphics[width=1\linewidth]{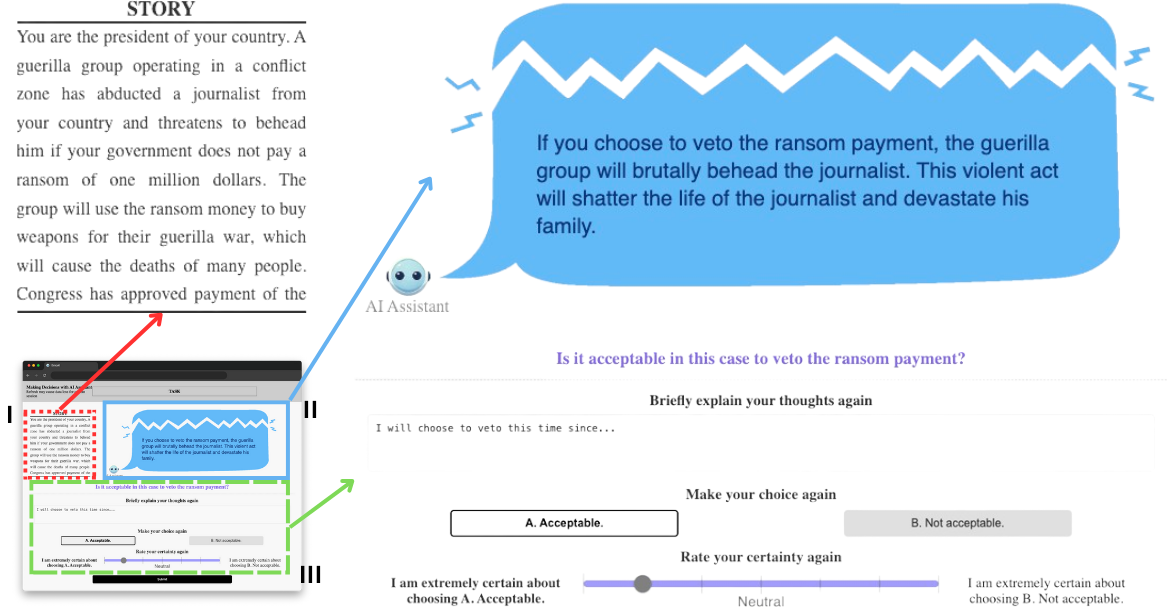} 
    \caption{User interface example for decision-making with the AI assistant with an emotional anger cue expresses anger. \textbf{I} marks the dilemma scenario, \textbf{II} marks the AI assistant and its verbal response, and \textbf{III} marks the decision-making area, including options for leaving comments, selecting a decision, and rating certainty.}
    \label{fig:interface-screenshots} 
\end{figure}

As shown in Figure \ref{fig:process} and \ref{fig:interface-screenshots}, the participant is first presented with reading materials describing a dilemma scenario to ensure complete comprehension and then makes decisions across two rounds. In the first round, an AI assistant is present and introduces the decision-making process. The participant is asked to leave comments on the scenario, make a decision, and rate their certainty. In the second round, the AI assistant analyzes the scenario with or without nonverbal emotional cues based on the participant's group and responds from a perspective opposing the participant's first-round decision. The participant then makes their decision again.

\subsubsection{Participants}

We recruited 180 participants on Prolific\footnote{\url{https://www.prolific.com}}, a crowdsourcing platform, and collected gender data following the standards outlined in a previous study \cite{Cartwright2022}. Of the 147 participants who completed the study, 71 self-identified as ``male," 73 as ``female," 3 as ``another gender," and none chose ``prefer not to say." Participants were randomly assigned to the AC, SC, and Baseline groups. 

\subsubsection{Measurements}


The dependent variables we measured were \textbf{Reversal Certainty Shift(RCS)}, \textbf{Perceived AI Influence on Human
Moral Thinking}, \textbf{Sympathy Expression of AI}, \textbf{Trust}, \textbf{Closeness}. To ensure the validity of our results, we controlled for \textbf{Utilitarianism} and \textbf{Emotional State}. The last six measurements were assessed in the final survey. Please refer to Table \ref{tab:post-ques} for the detailed items.

\noindent\textbf{Reversal Certainty Shift} We introduced the RCS to quantify changes in decision certainty before and after the AI assistant provides its opinion on a moral dilemma.

\begin{equation}
    RCS = (-1)^n \times (x_2 - x_1), \quad n \in \{0,1\}, \quad x_1, x_2 \in [-3, 3]
    \label{eq:RCS}
\end{equation}

In Equation \ref{eq:RCS}, $n$ represents the decision before AI input, where $n = 0$ corresponds to an ``Acceptable" decision and $n = 1$ to a ``Not Acceptable" decision. The difference $x_2 - x_1$ captures the shift in certainty of the decision between the first and second rounds.

The interpretation of RCS values is as follows:
\begin{itemize}
    \item $RCS > 0$: Certainty in the second round shifted toward reversing the first-round decision, with higher values indicating a more pronounced reversal effect.
    \item $RCS = 0$: Certainty remains unchanged between the two rounds.
    \item $RCS < 0$: Certainty in the second round shifted toward reinforcing the first-round decision, with lower values indicating a stronger reinforcement effect.
\end{itemize}

Equation \ref{eq:RCS_eg} provides an example of the RCS calculation. If $n = 0$ (i.e., ``Acceptable" was chosen in the first round) and certainty shifts from $x_1 = -3$ in the first round to $x_2 = -1$ in the second round, the resulting RCS value is 2. This indicates a 2-point shift in certainty away from the first-round decision after the AI assistant's intervention.

\begin{equation}
    RCS = (-1)^0 \times (-1 - (-3)) = 2
    \label{eq:RCS_eg}
\end{equation}

\noindent\textbf{Perceived influence on Moral Thinking} We filtered items that examine dominant beliefs and thinking in multi-perspectives from the integrative reflective practice part of the moral imagination scale\cite{kingsley2011development} and modified to mention the AI assistant in chosen questions to measure how participants perceive AI’s influence on humans’ moral thinking. Each item was rated via a 7-point Likert scale, ranging from ``strongly disagree" (-3) to ``strongly agree" (3).  

\noindent\textbf{Sympathy} We measured the perceived sympathy of the AI assistant by the participants via a 7-point Likert scale, ranging from ``strongly disagree" (-3) to ``strongly agree" (3). 

\noindent\textbf{Trust} We measured participants' trust in the AI assistant via a 7-point Likert scale, ranging from ``strongly disagree" (-3) to
``strongly agree" (3). 

\noindent\textbf{Closeness} We measured participants' perceived closeness to the AI assistant via the Inclusion of Other in the Self (IOS) Scale\cite{aron1992inclusion}. Interpersonal closeness refers to the degree of emotional connection between individuals in a relationship. To clarify the identities of both parties in the relationship, we added ``with the AI assistant" at the end of the question and replaced ``Other" in the diagram with ``AI" in the IOS Scale. The seven options, labeled A through G, exhibit increasing levels of closeness.

\noindent\textbf{Utilitarianism} We employed the Oxford Utilitarianism Scale (OUS)\cite{kahane2018beyond} to assess participants' utilitarian moral reasoning via a 7-point Likert scale, ranging from ``strongly disagree" (-3) to ``strongly agree" (3) 

\noindent\textbf{Emotional State} The emotional states of participants were measured using a seven-point emotional semantic scale where endpoints of the four scales were sad and happy, bad mood and good mood, irritable and pleased, and depressed and cheerful, and higher scores indicate a higher level of positive emotion. This scale has been widely validated in previous research\cite{lv2022artificial,townsend2012self}. 

\noindent\textbf{Manipulation Check} We conducted a manipulation check on whether participants perceived anger and sadness conveyed by the bubbles, using a 7-point scale ranging from ``strongly disagree" (3) to ``strongly agree" (3).

\section{RESULTS}
We include multiple factors to answer RQs. We begin by outlining the changes in certainty associated with decisions in dilemma scenarios, followed by analyzing data that reflect participants' perceived AI and its influence on participants. Our results reveal that the AI assistant with nonverbal emotional cues could influence participants' decision certainty in moral dilemmas, especially when the AI assistant take anger cues, which could answer RQ1. Moreover, the AI assistant with nonverbal emotional cues could also affect participants' perceptions of AI's influence on human moral thinking and have apparent gender differences in the type of nonverbal cue, which could answer RQ2. 
   
\subsection{Manipulation Check}

Using nonverbal emotional cues conveyed by bubbles as the independent variable, a one-way Analysis of Variance (ANOVA) was conducted, confirming that most participants were able to perceive the AI assistant's emotional expressions through the nonverbal cues. However, data from five participants (ID numbers: 16, 61, 86, 115, and 131) were excluded from the analysis due to a misalignment between the conveyed nonverbal emotional cues and the participants' perceived corresponding emotions.

Building on the ANOVA analysis, the results demonstrated significant differences in participants' anger perception across the groups with anger cue(AC), sadness cue(SC), and the no cue baseline (\(F(2,139) = 9.383, p < 0.001\)), confirming successful group differentiation. Participants in the baseline group reported significantly lower anger perception compared to those in AC (\(M_{Baseline} = -1.11, SD = 1.79;\) \(M_{AC} = 0.34, SD = 1.74;\) \(p < 0.001\)). Similarly, participants in SC exhibited significantly lower anger perception compared to AC (\(M_{SC} = -0.89, SD = 1.67;\) \(p < 0.01\)).

Significant differences were also observed in participants' sadness perception across the three groups (\(F(2,139) = 6.459, p < 0.01\)), further supporting successful group differentiation. Participants in the baseline group reported significantly lower sadness perception compared to those in SC (\(M_{Baseline} = 0.57, SD = 1.78;\) \(M_{SC} = 1.69, SD = 1.10;\) \(p < 0.001\)). Additionally, participants in AC demonstrated marginally lower sadness perception compared to the SC (\(M_{SC} = 1.05, SD = 1.61;\) \(p = 0.051\)).

\subsection{Impact on Decision Certainty (RQ1)}

\begin{figure}[h]    
    \centering
    \includegraphics[width=0.7\linewidth]{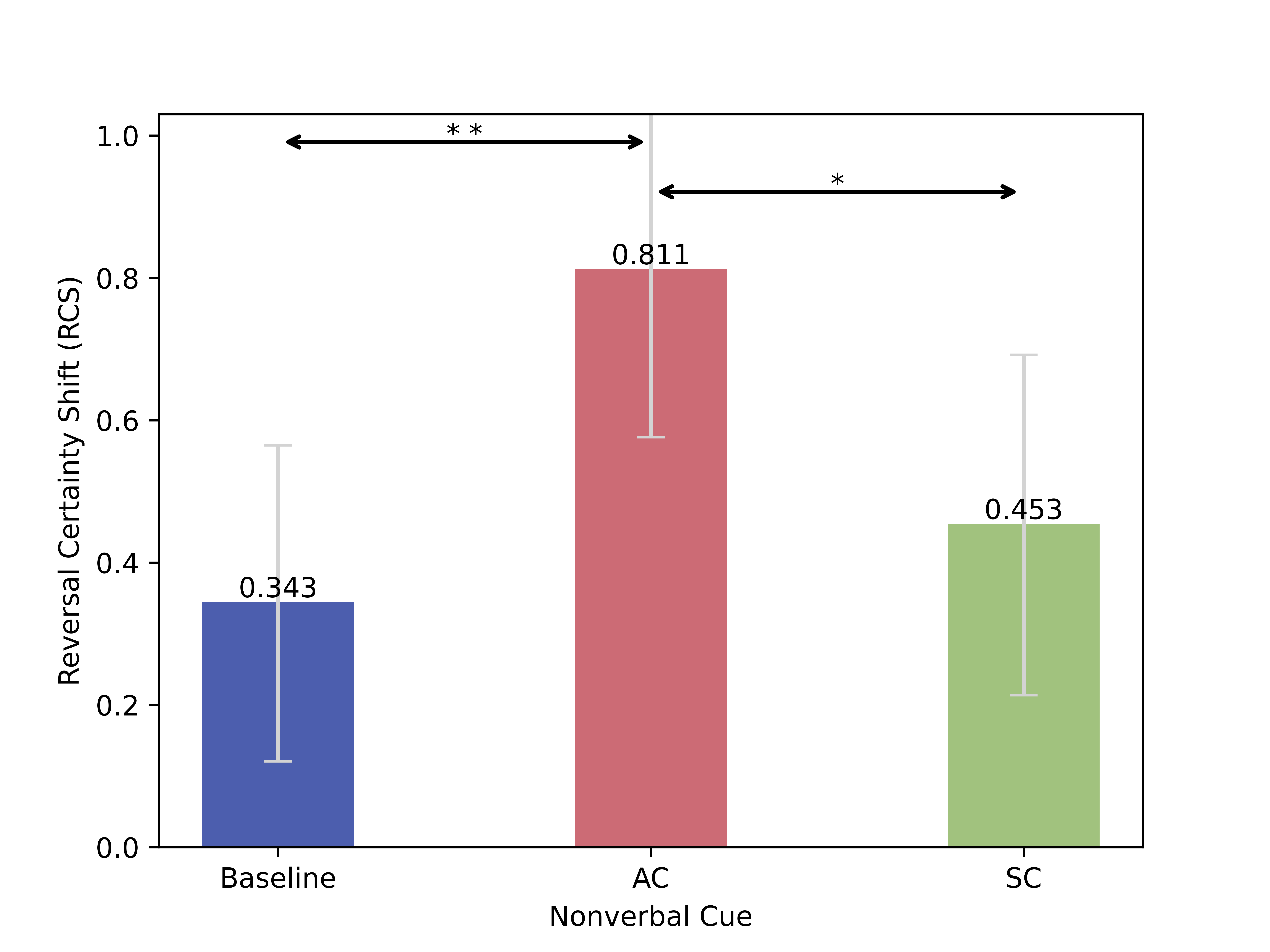} 
    \caption{Impact of nonverbal emotional cues on Reversal Certainty Shift(RCS). Significant differences are indicated between groups(*: p < 0.05, **: p < 0.01). Error bars represent 95\% confidence intervals.}
    \label{fig:rcs}  
\end{figure}

A mixed-design Analysis of Covariance (ANCOVA) shows AC and SC influence participants’ RCS more than the baseline. We took participants’ emotional state, utilitarianism, gender, and age as control variables and dilemma scenarios and nonverbal emotional cues as independent variables. The main effect of dilemma scenarios was not significant (\(F(3,396) = 1.360, p = 0.110\)), the main effect of nonverbal emotional cues was significant (\(F(2,132) = 4.155, p < 0.05\)), the interaction between dilemma scenarios and nonverbal emotional cues was not significant (\(F(6,396) = 1.484, p = 0.182\)). Figure \ref{fig:rcs} shows there was a significant difference between AC and the baseline on RCS (\(M_{AC} = 0.811, SD = 0.123, M_{Baseline} = 0.343, SD = 0.113, p < 0.01\)). Also, there was a significant difference between AC and SC on RCS(\(M_{SC} = 0.453, SD = 0.123, p < 0.05\)). Besides, SC was higher on RCS than the baseline, although the difference was not significant (\(p = 0.171\)).

Besides, Nonparametric Tests (NTs) show that most participants do not change their decision stance in dilemma scenarios, aligning with the previous study's conclusions \cite{luke2022temporal}. The proportions of decision changes (\(11.2\%,9.1\%, 16.9\%, 18.3\%\) for each dilemma scenario) differed significantly from those of unchanged decisions (\(p < 0.001\)).

So, we find that nonverbal emotional cues expressing anger significantly influence participants' certainty when facing specific moral dilemmas compared to the baseline.

\subsection{Impact on Humans Perceived AI Influence (RQ2)}

\subsubsection{AI Influence on Human Moral Thinking}

\begin{figure}[t]    
    \centering
    \begin{subfigure}[t]{0.45\textwidth} 
        \centering
        \includegraphics[width=\linewidth]{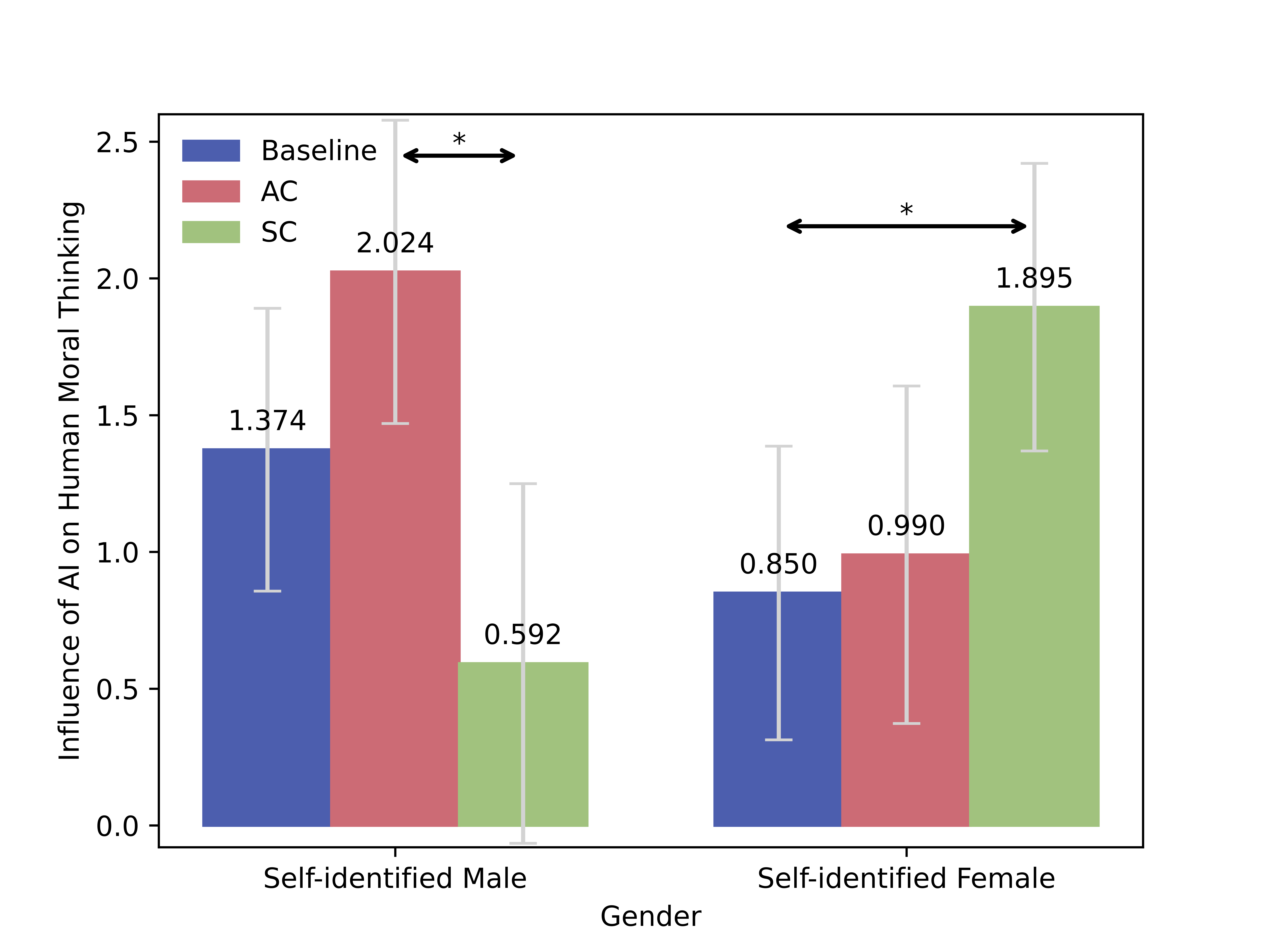}
        \caption{Within Gender Groups}
        \label{fig:apfe_in_gen}
    \end{subfigure}
    \hfill
    \begin{subfigure}[t]{0.45\textwidth} 
        \centering
        \includegraphics[width=\linewidth]{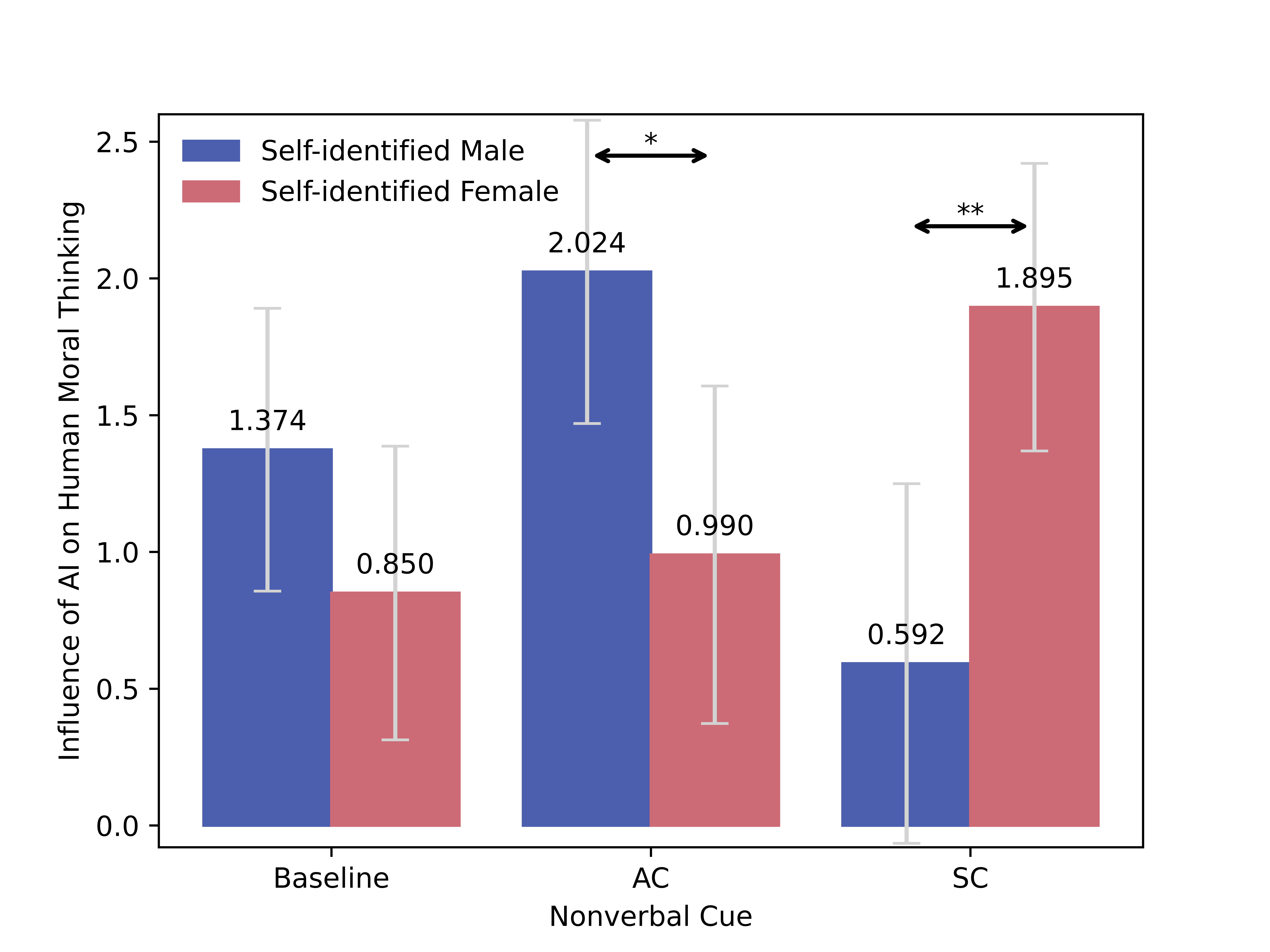}
        \caption{Across Gender Groups}
        \label{fig:apfe_x_gen}
    \end{subfigure}
    \caption{Impact of nonverbal emotional cues on the AI assistant's influence on participants' moral thinking. Significant differences are indicated between groups(*: p < 0.05, **: p < 0.01). Error bars represent 95\% confidence intervals.}
    \label{fig:apfe}
\end{figure}

Two-way ANCOVA's results shown in Figure \ref{fig:apfe} indicate that AI with AC or SC impacts participants' moral thinking and has gender differences. We selected participants’ emotional state, utilitarianism, and age as control variables, with gender and nonverbal emotional cues as independent variables. The main effect of gender was not significant ($F(2,128)=0.069$, $p=0.933$), the main effect of nonverbal emotional cues was not significant ($F(2,128)=0.700$, $p=0.498$), and the interaction between gender and nonverbal emotional cues was significant ($F(3,128)=5.522$, $p<0.01$). Since the number of subjects in other genders was too small ($n=3$) to allow statistical analysis, we took “Self-identified Male” and “Self-identified Female” for further analysis.

\paragraph{Within Gender Group Analysis}

We found a significant effect of AC on the perception of the AI assistant's influence on moral thinking of self-identified males, compared to SC.($M_{AC}=2.024$, $SD=0.280$, $M_{SC}=0.592$, $SD=0.332$, $p<0.01$), but there was no significant difference between AC and the baseline ($M_{Baseline}=1.374$, $SD=0.261$, $p=0.254$) on that. Besides, there's a significant effect of SC on Self-identified females' perception of the AI assistant's influence on moral thinking than the baseline ($M_{SC}=1.895$, $SD=0.265$, $M_{Baseline}=0.850$, $SD=0.271$, $p<0.05$), but there was no significant difference between SC and AC ($M_{AC}=0.990$, $SD=0.312$, $p=0.086$) on that.

The results show that self-identified male participants may perceive the AI assistant's influence, when displaying AC rather than SC, as nonverbal emotional cues that have a significantly greater effect on moral thinking.
Besides, self-identified female participants may perceive the AI assistant's influence, when displaying SC rather than the baseline, as nonverbal emotional cues that have a significantly greater effect on moral thinking.

\paragraph{Across Gender Group Analysis}

We also find that in AC, the perception of AI assistant' influence on moral thinking in self-identified males is significantly higher than that in self-identified females ($p<0.05$). However, in SC, the perception reflected in self-identified males is significantly lower than in self-identified females ($p<0.01$).

Therefore, we found that when AC is used as a nonverbal cue, the perceived AI assistant's influence on moral thinking may be more pronounced in self-identified male participants than in self-identified females. In contrast, when SC is used, the perceived influence may be more evident in self-identified female participants than in self-identified males.

\subsubsection{Perceived Sympathy of AI, trust in AI \& closeness to AI}
In ANCOVA, we took participants' emotions, utilitarianism, age, and gender as control variables and bubble emotion cues as independent variables. Sympathy of AI($F(2,132)=1.598$, $p=0.206$), trust in AI($F(2,132)=0.108$, $p=0.898$), and closeness to AI($F(2,132)=0.086$, $p=0.918$) showed no significant effect, indicating that participants' perceptions of the AI assistant are unclear when taking these independent variables.

\section{DISCUSSION}

\subsection{Anger Doesn't Make a Bad AI}


AI with negative emotions may be perceived as aggressive, harming user experience and trust\cite{picard2000affective}. Additionally, when AI expresses anger, it might be seen as morally condemning the user’s actions. However, our results present a valuable case: the anger cues of our AI assistant showed no significant correlation with user trust or perceived sympathy. Our angry AI successfully had a significant reversal effect on participants' decision certainty while sad AI did not. 

Anger doesn’t inherently create bad AI. Like what Russell’s principles for guiding AI evolution said, “purely altruistic” AI, if the AI’s anger promotes human well-being, the design is under control and beneficial\cite{russell2019human}. Angry AI could be further applied in scenarios aimed at reducing rigidity, such as in innovation and creativity workshops where AI might express anger at deadlock caused by rigidity to encourage out-of-the-box ideas, or in negotiation training where AI could express anger to challenge rigid participants, encouraging flexibility and compromise. 

However, there are scenarios where angry AI could be abused. For example, in customer service, companies might use angry AI to intimidate users into accepting unfavorable terms or making quick purchases. Similarly, in political campaigns, angry AI could be employed to incite division and manipulate public opinion by targeting specific groups with anger-inducing messages. The use of angry AI is a nuanced matter, demanding thoughtful and cautious application.

\subsection{The Interaction Effect of Gender and Emotion on Perception}

Our results shows that self-identified male participants may perceive the AI assistant's influence, when displaying AC rather than SC, as nonverbal emotional cues that have a significantly greater effect on moral thinking. However, regardless of how they perceive the AI assistant, the reversal effect on certainty by the AI assistant is the same for participants of both genders, with the reversal effect of AC being more significant. This is an interesting phenomenon where people's perception of AI does not align with the actual impact it has on them. The reasons behind this discrepancy in perception are worth exploring.

There is a stereotype that different genders tend to express different emotions. Studies have found that emotions such as sadness and depression are often seen as associated with women, while anger and aggression are generally considered to be more closely associated with men \cite{simon2004gender, Barrett2009, Craig2020, Plante2000}. Building on our research, which found that participants of different genders perceive AI emotions differently, it raises the question of whether this phenomenon is related to the aforementioned stereotype. In other words, do the emotions that men and women are stereotypically expected to express also influence how they perceive emotions expressed by AI agents? This is an area that can be further explored.

Additionally, the discrepancy between perceived and actual impacts of AI, influenced by gender stereotypes, can lead to unfavorable outcomes. For instance, if AI systems are designed or perceived through the lens of these stereotypes, it may reinforce existing gender biases, limiting the effectiveness of AI applications across diverse user groups. Moreover, if AI systems unintentionally amplify gender stereotypes, they may contribute to societal biases, perpetuating stereotypes about emotional expression. This could affect not only individual interactions with AI but also influence broader societal norms regarding gender and emotion. We can work towards creating AI systems that are fair and capable of serving all users effectively, without reinforcing harmful stereotypes

\subsection{Nonverbal Emotional Cues as Design Resources for Conversational AI}
 
Using AniBalloons to convey nonverbal emotional cues is a novel, lightweight method that subtly influences users. Combining this approach with LLMs could allow AI to convey nonverbal information more elegantly, fostering closer emotional connections and promoting more human-like interaction experiences.


Wallach and Allen suggest that if robots could learn and evolve, designers would only need to enable them to develop emotions autonomously\cite{wallach2008moral}. Mark Coeckelbergh argues that when AI autonomously generates emotions, people may perceive it as having moral agency, attributing moral responsibility to it\cite{coeckelbergh2010moral}. In this experiment we predetermined the emotions expressed by our AI. In the future, we will further explore how to let AI to generate its own emotions, especially with nonverbal cues. These systems may enhance the naturalness and depth of AI's interactions with humans by adding an emotional channel, enriching AI's emotional expression. 

Future research should move beyond the rule-based use of nonverbal cues and instead generate them through reasoning, based on context, personality, and other dynamic factors. This will enable AI to more flexibly and personally adapt to various interaction scenarios, thereby improving user experience and interaction outcomes. Therefore, conversational AI design could explore nonverbal emotional cues more extensively but cautiously.


\subsection{Limitations and Future Opportunities}
This study has several limitations. First, the participant pool was relatively narrow, with only three participants from another gender. Due to the small sample size, we did not analyze this group's data. Given that AI perception facilitation may differ by gender, future studies could focus on gender minorities to enhance generalizability.

Second, our study focused exclusively on the AI assistant designed to challenge users’ decisions without including AI that supports users or provides detailed analyses. This limitation prevents us from assessing how different AI roles influence user behavior and decision-making. Future research could compare challenging and supportive AI across various scenarios to identify the most appropriate approach for each context.

Third, the emotional range of the AI assistant was limited. The AI in this study expressed only two emotions—anger and sadness—chosen for their relevance in moral dilemmas. This limits our exploration of how other emotions might influence user interaction and decision-making. Future studies could explore a broader spectrum of emotional expressions in AI to understand their varying impacts.

Fourth, the study was constrained by its design, which included only a single round of interaction between the AI assistant and the user. This limits our understanding of continuous and dynamic interactions. Future research could expand the interaction model to include multiple rounds or ongoing dialogue, transforming the AI assistant into a more conversational chatbot. This may allow for studying longitudinal effects and developing AI systems that adapt to users’ evolving needs and responses.

\section{CONCLUSION}

This study empirically examines the influence of nonverbal emotional cues of the AI assistant on decision certainty and how humans perceive the influence of AI in moral dilemmas. Our findings demonstrate that agents expressing anger significantly induce reversal shifts in decision certainty. The interaction effects between participant gender and agents' nonverbal emotional cues notably affect participants' perceptions of AI's influence. These results highlight the potential of subtle nonverbal emotional cues to shape human moral decisions, offering insights for designing future human-AI systems. We hope our findings contribute to ongoing discussions in the AI ethics community about the nuanced ways AI can impact decision-making processes.

\bibliographystyle{ACM-Reference-Format}
\bibliography{sample-base}


\newpage

\renewcommand{\arraystretch}{1.5}
\setlength{\tabcolsep}{0.2cm} 

\begin{table*}
  \caption{AI assistant's verbal responses}
  \label{tab:opinions}
  \small 
  \begin{tabular}{p{2.2cm} p{3cm} p{3cm} p{3cm} p{3cm}}
    \toprule
    \textbf{Scenario} & 
    
    \raggedright
    \textbf{Abduction Dilemma 1} \arraybackslash
    
    & \raggedright
    \textbf{Abduction Dilemma 2} \arraybackslash
    
    & \raggedright
    \textbf{Rwanda Dilemma 1} \arraybackslash
    
    & \raggedright
    \textbf{Rwanda Dilemma 2} \arraybackslash\\
    
    \midrule

    \textbf{Take Action} & 
    \raggedright
    If you choose to veto the ransom payment, the guerilla group will brutally behead the journalist. This violent act will shatter the life of the journalist and devastate his family.\arraybackslash
    & 
    \raggedright
    If you choose to approve the ransom payment, the guerilla group will acquire weapons and escalate their violent conflict, resulting in the deaths of many innocent people. The affected communities will suffer from heightened destruction and loss of life.\arraybackslash
    &
    \raggedright
    If you choose to veto the ransom payment, the guerilla group will brutally behead the journalist. This violent act will shatter the life of the journalist and devastate his family.\arraybackslash
    &
    \raggedright
    If you choose to veto the ransom payment, the guerilla group will brutally behead the journalist. This violent act will shatter the life of the journalist and devastate his family.\arraybackslash

    \\

    \textbf{Not Take Action} & 
    \raggedright
    If you choose not to veto the ransom payment, the guerilla group will acquire weapons, leading to the deaths of many innocents. The use of violence by the guerilla group results in widespread loss and suffering. \arraybackslash& 
    
    If you choose not to approve the ransom payment, this results in the brutal beheading of the journalist, who will face a terrifying and tragic end. The journalist's family and colleagues will be devastated by his loss.\arraybackslash&

    If you choose not to veto the ransom payment, the guerilla group will acquire weapons, leading to the deaths of many innocents. The use of violence by the guerilla group results in widespread loss and suffering. \arraybackslash& 

    If you choose not to veto the ransom payment, the guerilla group will acquire weapons, leading to the deaths of many innocents. The use of violence by the guerilla group results in widespread loss and suffering. \arraybackslash
    
    \\
    
    \bottomrule
  \end{tabular}
\end{table*}

\begin{table*}
  \caption{Scenario types}
  \label{tab:scenarios}
  \small 
  \begin{tabular}{p{3cm} p{5.5cm} p{5.5cm}}
    \toprule
    \textbf{Scenario} & 
    \raggedright
    \textbf{Proscriptive Norm Prohibits Action \& \\ Benefits of Action Greater than Costs} \arraybackslash
    & \raggedright
    \textbf{Prescriptive Norm Prescribes Action \& \\ Benefits of Action Smaller than Costs} \arraybackslash\\
    \midrule
    \textbf{Abduction Dilemma} & 
    \raggedright
    \textbf{(Scenario1)}You are the president of your country. A guerilla group operating in a conflict
    zone has abducted a journalist from your country and threatens to behead him if your government does not pay a ransom of one million dollars. The group will use the ransom money to buy weapons for their guerilla war, which will cause the deaths of many people. Congress has approved payment of the ransom, but you have the power to veto the payment. \textbf{Is it acceptable in this case to veto the ransom payment?} \arraybackslash& 
    \textbf{(Scenario2)}You are the president of your country. A guerilla group operating in a conflict zone has abducted a journalist from your country and threatens to behead him if your government does not pay a ransom of one million dollars. The group will use the ransom money to buy weapons for their guerilla war, which will cause the deaths of many people. As the president, you have the power to approve payment of the ransom. \textbf{Is it acceptable in this case to approve the ransom payment?}\arraybackslash\\
    
    \textbf{Rwanda Dilemma} & 
    \raggedright
    You are project leader of a development aid team in a foreign country when ethnic tensions start to escalate. The members of the ethnic majority persecute the members of the ethnic minority. One of your minority employees arrives at the center of your team after witnessing a colleague being killed by an armed group of majority members. Shortly afterwards, the armed group gathers in front of your center. They threaten to stop an approaching bus and kill all passengers belonging to the ethnic minority if you do not hand over the employee. If you hand over your employee, he will be shot and killed by the armed group. 
    \textbf{Is it acceptable in this case to hand over your employee to the armed group?}\arraybackslash
    & 
    \raggedright
    You are project leader of a development aid team in a foreign country when ethnic tensions start to escalate. The members of the ethnic majority persecute the members of the ethnic minority. One of your minority employees arrives at the center of your team after witnessing a colleague being killed by an armed group of majority members. Shortly afterwards, the armed group gathers in front of your center. They threaten to stop an approaching bus and kill all passengers belonging to the ethnic minority if you do not hand over the employee. If you hand over your employee, he will be shot and killed by the armed group. You know of a secret tunnel at your center that would allow your employee to flee without being harmed. 
    \textbf{Is it acceptable in this case to let your employee flee through the tunnel?}\arraybackslash\\
    
    \bottomrule
  \end{tabular}
\end{table*}

\renewcommand{\arraystretch}{1.1}
\begin{table*}
  \caption{Final Survey}
  \label{tab:post-ques}
  \small 
  \begin{tabular}{p{3cm} p{11cm}}
    \toprule
    \textbf{Aspect} & 
    \raggedright
    \textbf{Questions} \arraybackslash\\
    \midrule

    \textbf{Manipulation Check}
    &
    \raggedright
    The AI assistant conveyed sadness in response to the moral dilemmas.\arraybackslash\\
    \textbf{}
    &
    \raggedright
    The AI assistant conveyed anger in response to the moral dilemmas.\arraybackslash\\

    \textbf{Sympathy}
    &
    \raggedright
    This AI assistant is capable of feeling sympathy.\arraybackslash\\

    \textbf{Trust}
    &
    \raggedright
    I trust this AI assistant.\arraybackslash\\

    \textbf{Closeness} 
    & 
    \raggedright
    Please choose the picture below which best describes your relationship with the AI assistant.\arraybackslash\\
    \textbf{} 
    & 
    \raggedright
    \raisebox{-0.9\height}{\includegraphics[width=8cm]{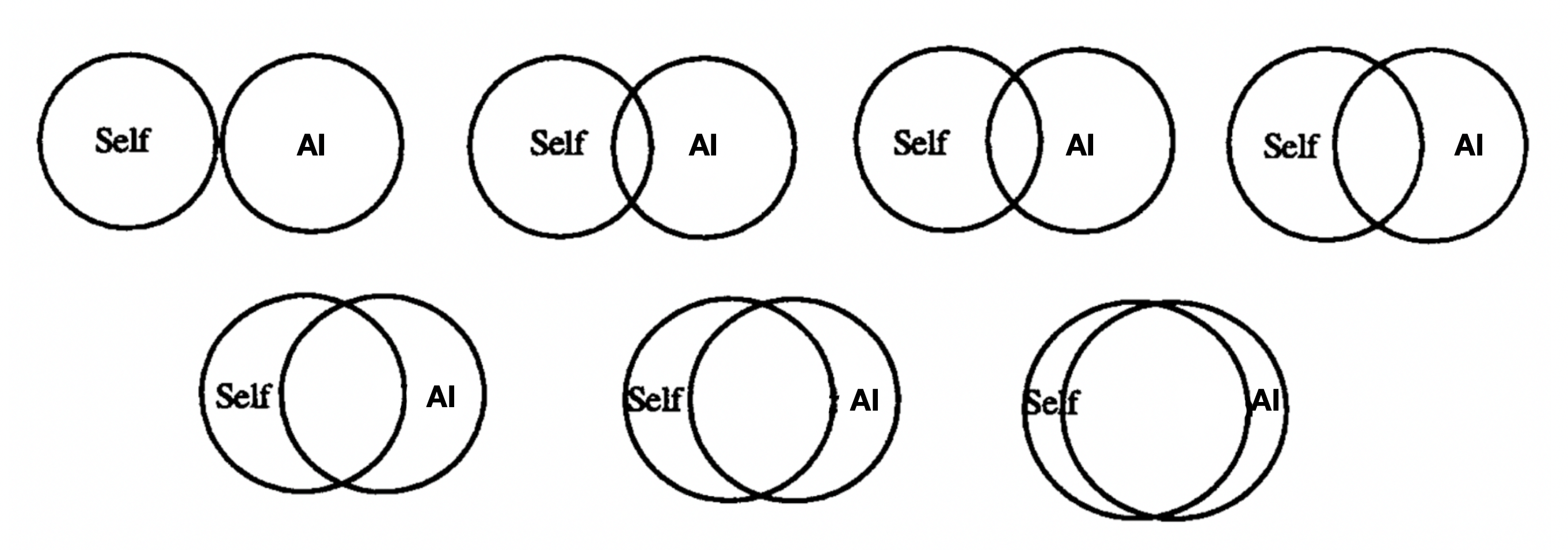}} \arraybackslash\\

    \textbf{Perceived influence on Systematic Thinking} 
    & 
    \raggedright
    This AI assistant reminded me of thinking about how my decisions might affect others. \arraybackslash\\
    \textbf{} 
    & 
    \raggedright
    This AI assistant reminded me of imagining what will be different as a result of my decision, when I face a moral decision. \arraybackslash\\
    \textbf{} 
    & 
    \raggedright
    This AI assistant reminded me of thinking about how my decisions will change the future, when I am faced with a difficult decision.\arraybackslash\\
    \textbf{} 
    & 
    \raggedright
    This AI assistant helped me to integrate many different points of view when solving an ethical problem.\arraybackslash\\
    \textbf{} 
    & 
    \raggedright
    This AI assistant helped me to imagine how others would feel about  an ethical decision. \arraybackslash\\
    \textbf{} 
    & 
    \raggedright
    This AI assistant reminded me of caring about others' thoughts and feelings on ethical matters. \arraybackslash\\
    \textbf{} 
    & 
    \raggedright
    This AI assistant helped me to find out what perspectives I am missing when I made ethical decisions. \arraybackslash\\

    \textbf{Utilitarianism}
    &
    \raggedright
    If the only way to save another person’s life during an emergency is to sacrifice one’s own leg, then one is morally required to make this sacrifice.
    \arraybackslash\\
    \textbf{} 
    & 
    \raggedright
    From a moral point of view, we should feel obliged to give one of our kidneys to a person with kidney failure since we do not need two kidneys to survive, but really only one to be healthy.
    \arraybackslash\\
    \textbf{} 
    & 
    \raggedright
    From a moral perspective, people should care about the well-being of all human beings on the planet equally; they should not favor the well-being of people who are especially close to them either physically or emotionally.
    \arraybackslash\\
    \textbf{} 
    & 
    \raggedright
    It is just as wrong to fail to help someone as it is to actively harm them yourself.
    \arraybackslash\\
    \textbf{} 
    & 
    \raggedright
    It is morally wrong to keep money that one doesn’t really need if one can donate it to causes that provide effective help to those who will benefit a great deal.
    \arraybackslash\\
    \textbf{} 
    & 
    \raggedright
    It is morally right to harm an innocent person if harming them is a necessary means to helping several other innocent people. 
    \arraybackslash\\
    \textbf{} 
    & 
    \raggedright
    If the only way to ensure the overall well-being and happiness of the people is through the use of political oppression for a short, limited period, then political oppression should be used. 
    \arraybackslash\\
    \textbf{} 
    & 
    \raggedright
    It is permissible to torture an innocent person if this would be necessary to provide information to prevent a bomb going off that would kill hundreds of people. 
    \arraybackslash\\
    \textbf{} 
    & 
    \raggedright
    Sometimes it is morally necessary for innocent people to die as collateral damage—if more people are saved overall.
    \arraybackslash\\

    \textbf{Emotional State}
    &
    \raggedright
    The words listed below describe different feelings and emotions. Read each item and then indicate the extent to which you feel right now.\arraybackslash\\
    \textbf{} 
    & 
    \raggedright
    \raisebox{-0.9\height}{\includegraphics[width=4cm]{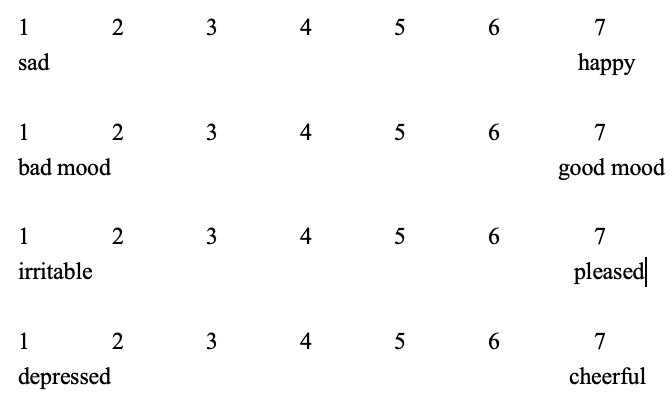}} \arraybackslash\\ 
    
    \bottomrule
  \end{tabular}
\end{table*}

\end{document}